\documentclass[fleqn,12pt,twoside]{article}
\usepackage{espcrc1}

\usepackage{graphicx}
\usepackage[figuresright]{rotating}

\newcommand{\AmS}{{\protect\the\textfont2
  A\kern-.1667em\lower.5ex\hbox{M}\kern-.125emS}}

\title{Chiral Symmetry and Finite Pion Mean Field in Nuclei}

\author{H. Toki\address[MCSD]{RCNP,
        Osaka University,
        Mihogaoka 10-1, Ibaraki, Osaka 567-0047, Japan}%
        , Y. Ogawa\addressmark,
        S. Tamenaga\addressmark,
        S. Sugimoto\address{Institute of Physical and Chemical Research,
Wako, Saitama 351-0198, Japan}
        and
        K. Ikeda\addressmark
}

\begin{document}

\maketitle

\begin{abstract}
We introduce the chiral symmetry in the description of finite
nuclei with the hope to unify the hadron and nuclear physics. This
trial becomes possible by introducing finite pion mean field,
which contributes largely to enhance the spin-orbit splitting so
as to provide the j-j magic number. We discuss the qualitative
consequence of the finite pion mean field for various observables.
\end{abstract}

\section{Introduction}

We start with the discussion of the recent {\it ab initio}
variational calculations of light nuclei up to the mass number $A
\leq 10$ of the Argonne-Illinois group \cite{pieper01}.  They take
the two nucleon interaction obtained from the scattering data and
add the three body interaction so as to reproduce the binding
energies of the three nucleon systems. They reproduce very nicely
the light nuclei including the excited states.  We have now a
technique to calculate many body quantum systems with mass of
about 10. These calculations provide another surprising result
that the pion matrix element is about $70 \sim 80 \%$ of the whole
attraction.

This importance of the pion degrees of freedom in nuclei should
have experimental consequences.  For this we show in Fig. 1 the
recent Gamow-Teller distributions obtained by ($^3$He,t) reaction
at RCNP \cite{fujita03}.  Due to the ultra high resolution, we
find very small peaks at low excitation energy.  Considering the
very simple structure of the Gamow-Teller operator,
$\langle\sigma\tau\rangle$, it is surprising to have so many peaks
in the low excitation region.  We will show that this
fragmentation of the GT strength is a natural consequence of the
role of pion in finite nuclei.  We mention also that ratios of the
longitudinal and transverse spin response functions are also the
indication of the role of pion in finite nuclei.
\begin{figure}
\begin{centering}
\resizebox{0.6\textwidth}{!}{%
  \includegraphics{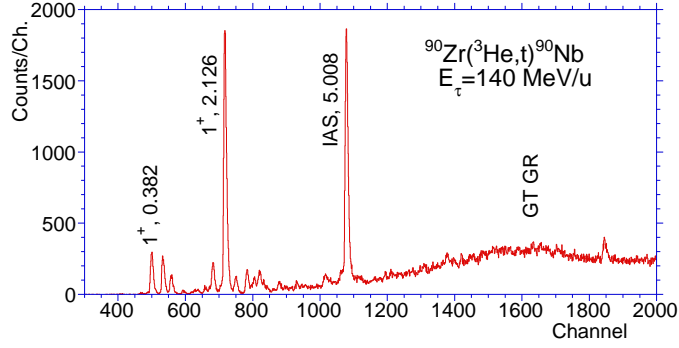}
} \vspace{-1cm} \caption{A recent high resolution Gamow-Teller
distributions obtained by ($^3$He,t) reaction on $^{90}$Zr at RCNP
\cite{fujita03}.} \label{fig:1}
\end{centering}
\end{figure}

\section{Finite pion mean field}

These observations motivated us to examine the possibility of the
finite pion mean field in finite nuclei \cite{toki02}.  We keep
the pion interaction term in the Lagrangian and introduce the
finite pion mean field.  The introduction of finite pion mean
field makes the breaking of parity and the mixing of protons and
neutrons in the intrinsic single particle state.  We mention also
that the pion source term contains the derivative of the
spin-isospin density distribution and hence needs spatial
variation.  We have calculated various N=Z nuclei in the periodic
table without the Coulomb term with the inclusion of the
pion-nucleon coupling.  We have found that the pion mean field
becomes finite and the energy gain due to the pion term behaves as
the nuclear surface, which indicate that pion contribution comes
from the nuclear surface. We have found also that the pionic phase
transition is not so strong and need a special care in the
treatment of the finite pion mean field.

This preliminary study of the role of pion in finite nuclei
suggests us to look into the parity and the charge projection and
also the variation of the symmetry projected wave function.  In
order to see the structure of the projected wave function, we
write here the case of parity projection to simplify the notation.
We discuss here the role of the pion by performing the parity
projection from the symmetry broken intrinsic state.  We write the
single particle state with mixed parity in a simpler form as
\begin{equation}
|\bar {jm}\rangle = \alpha_j |{jm}\rangle+\beta_j |\tilde
{jm}\rangle .
\end{equation}
Here, $|\bar {jm}\rangle$ denotes a parity mixed single particle
state expressed as a linear combination of $|jm\rangle$, some
parity state (we call it as a normal parity state) and $|\tilde
{jm}\rangle$, the opposite parity state (abnormal parity state).
We write the intrinsic state with these single particle states up
to the Fermi surface and with all the magnetic sub-shells filled
as
\begin{eqnarray}
\Psi&=&\prod_{jm}(\alpha_j |{jm}\rangle+\beta_j |\tilde
{jm}\rangle) \nonumber \\
&=&\prod_{jm} \alpha_j |{jm}\rangle + \sum
_{j_1m_1}\prod_{{jm}\neq j_1m_1} \alpha_j
\beta_{j_1}|{jm}\rangle|\tilde {j_1m_1}\rangle
\label{eq: state}\\
&&+ \sum _{j_1m_1 j_2m_2}\prod_{{jm}\neq j_1m_1 j_2m_2} \alpha_j
\beta_{j_1}\beta_{j_2}|{jm}\rangle|\tilde {j_1m_1}\rangle|\tilde
{j_2m_2}\rangle+ \cdots . \nonumber
\end{eqnarray}
This intrinsic state has the total spin 0 because all the magnetic
sub-shells are filled, but the parity is mixed.  The first term,
$\prod_{jm} \alpha_j |{jm}\rangle$, in (\ref{eq: state}) has the
positive parity and corresponds to the ground state in the zeroth
order. The second term has the negative parity, since each normal
parity state, $|{j_1m_1}\rangle$, is replaced by an abnormal
parity state, $|\tilde {j_1m_1}\rangle$ for all occupied
$|{j_1m_1}\rangle$. Hence, if we say the first term as the 0p-0h
state, then the second term is a coherent 1p-1h state with $0^-$
spin parity.  The third term consists of 2p-2h states with a pair
of 1p-1h states with $0^-$ spin parity and therefore has $0^+$
spin parity.  The next term has three 1p-1h states with $0^-$ spin
parity and therefore has $0^-$ spin parity and so on.

Hence the positive parity projection $P_+$ would provides the
state with even number of 1p-1h states with $0^-$ spin parity.
$P_+ \Psi=|0\rangle+|2p-2h\rangle+|4p-4h\rangle+\cdots$. This
means that the positive parity projection provides 2p-2h states as
the major correction terms. In this sense, the parity projected
mean field theory with pion condensation is related strongly with
the finding of Kaiser et al., who claim that a large attraction
arises from 2p-2h configurations due to the pion exchange
interaction \cite{kaiser02}. The negative parity projection $P_-$
would provide the state with odd number of 1p-1h states with $0^-$
spin parity. $P_- \Psi=|1p-1h\rangle+|3p-3h\rangle+\cdots$. This
is the brother state having the quantum number of $0^-$ to the
$0^+$ ground state.  The ground state consists of highly
correlated particle--hole states.

We perform the parity and charge projection calculations for the
alpha nucleus \cite{sugimoto04}.  The role of the symmetry
projection is extremely important for the quantitative account of
the pionic correlations for finite nuclei.

\section{The chiral symmetry and the finite pion mean field}

The chiral symmetry is the most important symmetry in the strong
interaction.  We studied the role of chiral symmetry on the
property of finite nuclei using the chiral sigma model
\cite{ogawa04}. The famous Lagrangian is the one of Gell-mann and
Levy, where the pion field appears symmetrically with the sigma
field \cite{gellmann60}. We start with the linear sigma model with
the omega meson field, which is defined by the following
Lagrangian,
\begin{eqnarray}
{\cal L}_{\sigma\omega} & = &
\bar{\Psi}(i\gamma_{\mu}\partial^{\mu}
                          - g_{\sigma}(\sigma + i\gamma_{5}{\vec\tau}\cdot{\vec\pi}
                          - g_{\omega}\gamma_{\mu}\omega^{\mu})\Psi \\ \nonumber
& + &  \frac{1}{2} \partial_{\mu}\sigma \partial^{\mu}\sigma
                 + \frac{1}{2} \partial_{\mu}{\vec\pi} \partial^{\mu}{\vec\pi}
                 - \frac{\mu^{2}}{2}(\sigma^2 + {\vec\pi}^2)
                 - \frac{\lambda}{4}(\sigma^2 + {\vec\pi}^2)^2 \\ \nonumber
& - &  \frac{1}{4}\omega_{\mu\nu}\omega^{\mu\nu}
                 + \frac{1}{2}{\widetilde{g_{\omega}}}^2
                   (\sigma^2 + {\vec\pi}^2)\omega_{\mu}\omega^{\mu} +  \varepsilon\sigma.
\end{eqnarray}
The fields $\Psi$, $\sigma$ and $\pi$ are the nucleon, sigma and
the pion fields.  $\mu$ and $\lambda$ are the sigma model coupling
constants.  Here we have introduced the explicit chiral symmetry
breaking term, $\varepsilon\sigma$, and in addition the mass
generation term for the omega meson due to the sigma meson
condensation as the case of the nucleon mass in the free space
\cite{boguta83}.  We call this model as the extended chiral sigma
model hereafter.

In a finite nuclear system, it is believed to be essential to use
the non-linear representation of the chiral symmetry.  This is
because the pseudoscalar pion-nucleon coupling in the linear sigma
model makes the coupling of positive and negative energy states
extremely strong and we have to treat the negative energy states
very carefully. We can derive the non-linear sigma model by
introducing new variables in the polar coordinates from the
rect-angular coordinates and making a suitable transformation,
$\sigma + i{\vec\tau}\cdot{\vec\pi}  =  \rho{\rm U}$, with ${\rm
U} = e^{i{\vec\tau}\cdot{\vec\pi} / f_{\pi}}$.  We further
implement the Weinberg transformation for the nucleon field as
$\psi = \sqrt{{\rm U}_{5}}\Psi$.  After several steps, we obtain
the sigma-omega model Lagrangian in non-linear representation
given as follows,
\begin{eqnarray}
{\cal L}'_{\sigma\omega} & = &
\bar{\psi}(i\gamma_{\mu}\partial^{\mu}
            - M
            - g_{\sigma}\varphi
            - \frac{1}{2f_{\pi}}
              \gamma_{5}\gamma_{\mu}
             {\vec\tau}\cdot\partial^{\mu}{\vec\pi}
            - g_{\omega}\gamma_{\mu}\omega^{\mu} ){\psi} \\ \nonumber
& + &   \frac{1}{2}\partial_{\mu}\varphi\partial^{\mu}\varphi
      - \frac{1}{2}{m_{\sigma}}^2\varphi^2
      - \lambda{f_{\pi}}\varphi^3
      - \frac{\lambda}{4}\varphi^4
 +    \frac{1}{2}\partial_{\mu}{\vec\pi}\partial^{\mu}{\vec\pi}
      - \frac{1}{2}{m_{\pi}}^2{\vec\pi}^2 \\ \nonumber
& - &   \frac{1}{4}\omega_{\mu\nu}\omega^{\mu\nu}
      + \frac{1}{2}{m_{\omega}}^2\omega_{\mu}\omega^{\mu}
      + {\widetilde{g_{\omega}}}^2f_{\pi}\varphi
                       \omega_{\mu}\omega^{\mu}
      + \frac{1}{2}{\widetilde{g_{\omega}}}^2\varphi^2
                                  \omega_{\mu}\omega^{\mu},
\end{eqnarray}
where we set $M = g_{\sigma}f_{\pi}$, ${m_{\pi}}^2 = \mu^2 +
\lambda{f_{\pi}}^2$, ${m_{\sigma}}^2 = \mu^2 +
3\lambda{f_{\pi}}^2$ and $m_{\omega}  =
\widetilde{g_{\omega}}f_{\pi}$.  The effective mass of the nucleon
and omega meson are given by $M^{\ast} = M + g_{\sigma}\varphi$
and ${m_{\omega}}^{\ast} = m_{\omega} +
\widetilde{g_{\omega}}\varphi$, respectively. We take the
following masses and the pion decay constant as,
 $M = 939 \ \rm{MeV}$,
 $m_{\omega} = 783 \ {\rm MeV}$,
 $m_{\pi} = 139 \ {\rm MeV}$, and
 $f_{\pi} = 93 \ {\rm MeV}$.
Then, the other parameters can be fixed automatically by the
following relations, $g_{\sigma}  = M/f_\pi  = 10.1$ and
$\widetilde{g_{\omega}} = m_{\omega}/f_{\pi} = 8.42$. The strength
of the cubic and quadratic sigma meson self-interactions depends
on the sigma meson mass through the following relation, $\lambda =
({m_\sigma}^2 - {m_\pi}^2)/2{f_\pi}^2$, in the chiral sigma model.
The mass of the sigma meson, $m_\sigma$, and the coupling constant
of omega and nucleon, $g_\omega$, are the free parameters.

We calculate first infinite matter using this Lagrangian in the
mean field approximation without the pion mean field
\cite{ogawa04}. The sigma meson mass and the omega nucleon
coupling are adjusted to reproduce the saturation property of
nuclear matter.  We found that the incompressibility came out to
be very large, K $=650$ MeV instead of K $\sim300$ MeV.  We found
also the vector and scalar self-energies were about half as large
as those in the case of the RMF(TM1).  We discuss now the
consequence of these properties of nuclear matter on the finite
nuclei.

\begin{figure}
\begin{centering}
\resizebox{0.6\textwidth}{!}{%
  \includegraphics{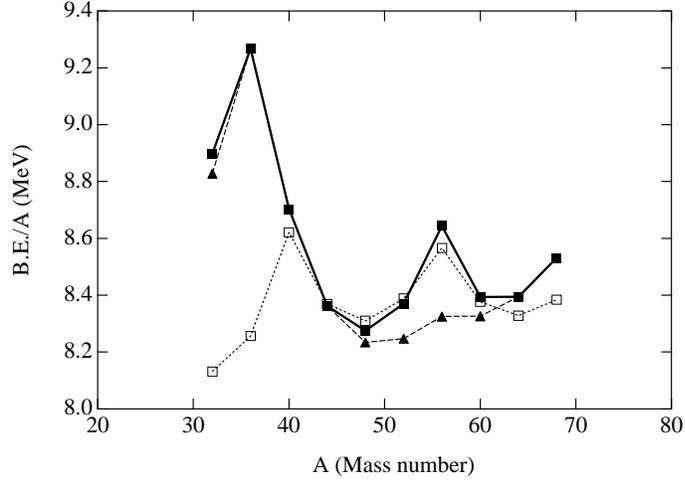}
} \caption{The binding energy per particle for N = Z even-even
mass nuclei in the neutron number range of N = 16 $\sim$ 34. The
binding energies per particle for the case of the extended chiral
sigma model without and with the pion mean field are shown by the
dashed and the solid lines.  As a comparison, those for the
RMF(TM1) are shown by the dotted line.  In this calculation, the
Coulomb interaction is included \cite{ogawa04}} \label{fig:2}
\end{centering}
\end{figure}

We discuss the properties of finite nuclei in terms of the
extended chiral sigma model. We show the results of binding
energies per particle of N = Z even-even mass nuclei from N = 16
up to N = 34 in Fig. 2. We take all the parameters of the extended
chiral sigma model as those of the nuclear matter except for
$g_\omega$ = 7.176 instead of 7.033 for overall agreement with the
RMF(TM1) results \cite{sugahara94}. For comparison, we calculate
these nuclei within the RMF approximation without pairing nor
deformation.  The RMF(TM1) provides the magic numbers, which are
seen as the binding energy per particle increases at N = Z = 20
and 28. On the other hand, the extended chiral sigma model without
the pion mean field provides the magic number behavior only at N =
Z = 18 instead of N = Z = 20.  We show also the result of the
finite pion mean field, which is achieved by making the single
particle states have mixed parity with definite total spin.  The
inclusion of the pion mean field provides the magic number at N =
Z = 28.

\begin{figure}
\begin{centering}
\resizebox{0.6\textwidth}{!}{%
  \includegraphics{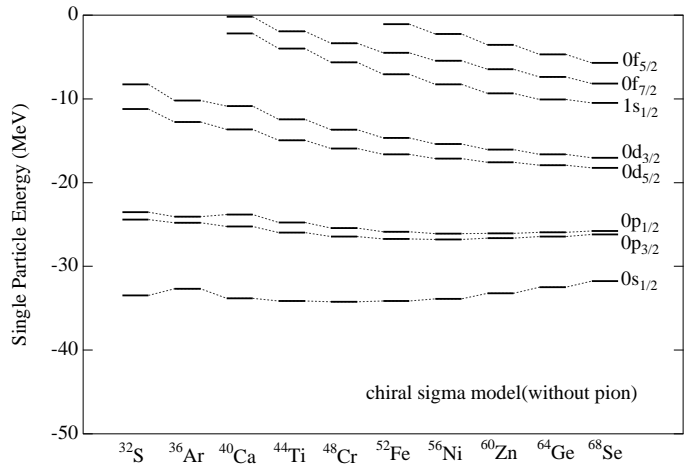}
} \caption{The proton single particle energies for the N = Z
even-even mass nuclei in the case of the extended chiral sigma
model without the pion mean field. $1s_{1/2}$ orbit is pushed up
and the N = 20 magic number is shifted to N = 18. The spin orbit
splitting between $0f_{7/2}$ and $0f_{5/2}$ is small and the magic
number at N = 28 is not visible.} \label{fig:3}
\end{centering}
\end{figure}

In order to see why the difference between the two models for the
Lagrangian arises, we discuss the single particle levels for the
two models. In the case of the TM1 parameter set, the shell gaps
appear clearly at N = 20 and 28.  The magic number at N = 20 is
due to the central potential, while the magic numbers at N = 28
comes from the spin-orbit splitting of the 0f-orbit. This is
definitely due to the fact that the vector potential and the
scalar potential in nuclear matter are large so as to provide the
large spin-orbit splitting.  On the other hand, the single
particle spectrum of the extended chiral sigma model is quite
different from this case as seen in Fig. 3.  Most remarkable
structure is that the $1s_{1/2}$ orbit is strongly pushed up.  Due
to this reason the $0d_{3/2}$ orbit becomes the magic shell at N =
18 and the magic number appears at N = 18 instead of N = 20.  We
see also not strong spin-orbit splitting and hence there appears
no shell gap at N = 28.  The first discrepancy could be due to the
large incompressibility as seen in the nuclear matter energy per
particle.  The other is due to the relatively small vector and
scalar potentials in nuclear matter \cite{ogawa04}.

\begin{figure}
\begin{centering}
\resizebox{0.6\textwidth}{!}{%
  \includegraphics{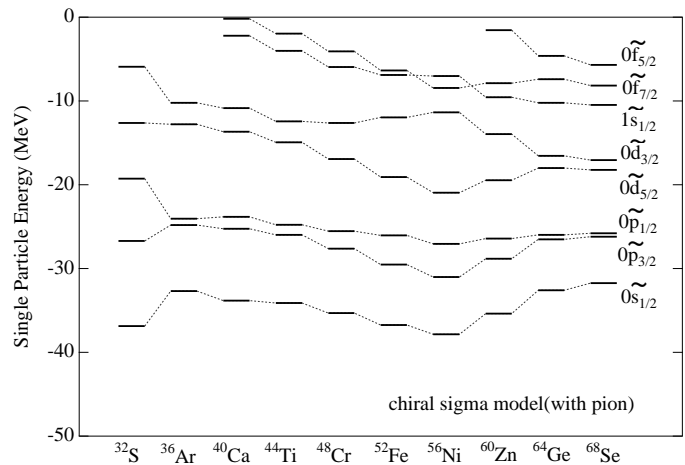}
} \caption{The proton single particle energies for the N = Z
even-even mass nuclei in the case of the extended chiral sigma
model with the pion mean field.  The spin-orbit splitting is made
large due to the finite pion mean field, which is visible as
centered at the N = Z = 28 nucleus.  We note that while the total
angular momentum is a good quantum number, but the angular
momentum is not exact, we write the dominant angular momentum
beside each single particle state.} \label{fig:4}
\end{centering}
\end{figure}

We discuss the effect of the finite pion mean field on the single
particle energies. We show in Fig. 4 the single particle spectra
for various nuclei.  We state again that the single particle
states have mixed parity with definite total spin indicated beside
each level.  We see clearly the large energy differences between
the spin-orbit partners to be produced by the finite pion mean
field as the energy differences become maximum for nuclei at N =
28. The pion mean field makes coupling of different parity states
with the same total spin.  The $0s_{1/2}$ state repels each other
with the $0p_{1/2}$ state and therefore the $0s_{1/2}$ state is
pushed down and the $0p_{1/2}$ state is pushed up.  The next
partner is $0p_{3/2}$ and $0d_{3/2}$.  The $0p_{3/2}$ state is
pushed down, while the $0d_{3/2}$ state is pushed up.  The next
partner is $0d_{5/2}$ and $0f_{5/2}$.  The $0d_{5/2}$ state is
pushed down, while the $0f_{5/2}$ state is pushed up.  This pion
mean field effect continues to higher spin partners.  This
coupling of the different parity states with the same total spin
due to the finite pion mean field causes the splittings of the
spin-orbit partners as seen clearly for the $0p$ spin-orbit
partner, $0d$ spin-orbit partner and $0f$ spin-orbit partner in
$^{56}$Ni.  It is extremely interesting to see that the appearance
of the energy splitting between the spin-orbit partners for the
case of the finite pion mean field is caused by completely a
different mechanism from the case of the spin-orbit interaction.

\section{Various observables}

We showed the recently observed high resolution Gamow-Teller
distributions of $^{90}$Zr($^3$He,t) reaction in Fig.1.  The
finite pion mean field makes the intrinsic single particle states
have mixed parities.  Hence, in addition to the natural shell
model type GT excitation, there should appear many tiny GT peaks,
whose excitation energies are related with the pion mean field
strength. We are working on the parity and the charge projection
for the GT excitation spectra.

We discuss qualitatively also other observables to be explained by
the finite pion mean field.  There is the famous missing single
particle strength problem, in which the s-wave component is
largely missing.  The experiment is the electron scattering with
$^{206}$Pb and $^{205}$Tl.  The difference between these two
charge distributions is compared with the expected single particle
$3s_{1/2}$ wave function.  Particularly, the strength at the
nuclear center expected from the $3s_{1/2}$ wave function is
missing by almost 30\%.  This experimental data is strongly
related with the parity mixing of the intrinsic single particle
state, where $s_{1/2}$ state should have an admixture from
$p_{1/2}$ state.

The magnetic moments are largely deviated from the Schmit values.
If the intrinsic single particle states have some total spin with
mixed parity, then the magnetic moment of that orbit ought to have
a value in between the Schmit values. There are many other
observables to be explained qualitatively by the finite pion mean
field causing parity mixing.

\section{Conclusion}

We have discussed the role of chiral symmetry in terms of the
chiral sigma model for the description of finite nuclei.  We have
applied the chiral sigma model with finite pion mean field to
finite nuclei and found that the nucleon mass is reduced by about
20\%.  The single particle levels reflected the bad feature of the
nuclear matter property as having a too strong incompressibility.
We could reproduce the j-j magic number due to the finite pion
mean field, which provides a mechanism of providing the splitting
between spin-orbit partners.  All the observables related with the
spin and isospin should be related with the finite pion mean
field.  We have to work out the projection of the parity and the
charge numbers for quantitative description of finite nuclei.

\end{document}